\documentclass{emulateapj}
\usepackage{natbib}
\usepackage{epsfig,color}
\usepackage{color}

\newcommand{\slowpokes}{SLoWPoKES}
\newcommand{\ha}{H\ensuremath{\alpha}}
\newcommand{\ewha}{\ensuremath{{\rm EW}~{{\rm H}{\alpha}}}}
\newcommand{\lha}{\ensuremath{{L}_{\rm H\alpha}/{L}_{\rm bol}}}

\shorttitle{H$\alpha$ Emission From Twin M Dwarfs}
\shortauthors{Gunning et al.}

\begin{document}

\title{H$\alpha$ Emission From Active Equal-mass, Wide M Dwarf Binaries$^{\dagger}$}

\author{
  Heather C. Gunning\altaffilmark{1,2}, 
  Sarah J. Schmidt\altaffilmark{1,3}, 
  James R. A. Davenport\altaffilmark{1}, 
  Saurav Dhital\altaffilmark{4,5}, 
  Suzanne L. Hawley\altaffilmark{1}, 
  Andrew A. West\altaffilmark{5}}
\altaffiltext{1}{Astronomy Department, Box 351580, University of Washington, Seattle, WA 98195, USA}
\altaffiltext{2}{Space Telescope Science Institute, Box 351580,3700 San Martin Drive, Baltimore, MD 21218, USA}
\altaffiltext{3}{Department of Astronomy, Ohio State University, 140 West 18th Avenue, Columbus, OH 43210, USA}
\altaffiltext{4}{Department of Physical Sciences, Embry-Riddle Aeronautical University, 600 South Clyde Morris Blvd., Daytona Beach, FL 32114, USA}
\altaffiltext{5}{Department of Astronomy, Boston University, 725 Commonwealth Avenue, Boston, MA 02215, USA}
\email{schmidt@astronomy.ohio-state.edu}
\footnotetext[$\dagger$]{This publication is partially based on observations obtained with the Apache Point Observatory 3.5-meter telescope, which is owned and operated by the Astrophysical Research Consortium.}

\begin{abstract}
We identify a sample of near-equal mass wide binary M dwarf systems from the {\slowpokes} catalog of common proper-motion binaries and obtain follow-up observations of their chromospheric activity as measured by the H$\alpha$ emission line. We present optical spectra for both components of 48 candidate M dwarf binaries, confirming their mid-M spectral types. Of those 48 coeval pairs, we find eight with H$\alpha$ emission from both components, three with weak emission in one component and no emission in the other, and 37 with two inactive components. We find that of the eleven pairs with at least one active component, only three follow the net trend of decreasing activity strength ({\lha}) with later spectral type. The difference in quiescent activity strength between the A and B components is larger than what would be expected based on the small differences in color (mass). For five binaries with two active components, we present 47 hours of time-resolved spectroscopy, observed on the ARC 3.5-m over twelve different nights. For four of the five pairs, the slightly redder (B) component exhibits a higher level of H$\alpha$ emission during the majority of the observations and the redder objects are the only components to flare. The full range of H$\alpha$ emission observed on these variable mid-M dwarfs is comparable to the scatter in H$\alpha$ emission found in single-epoch surveys of mid-M dwarfs, indicating that variability could be a major factor in the spread of observed activity strengths. We also find that variability is independent of both activity strength and spectral type. 
\end{abstract}


\section{Introduction}
\label{sec:intro}
M dwarfs are well known for harboring surface magnetic fields as
strong as several kG \citep[e.g.,][]{Johns-Krull1996,Valenti2001},
persisting on both sides of the fully convective boundary
\citep[$\sim$M4;][]{Chabrier1997}. The strength of the magnetic fields in M dwarfs may be intrinsically tied
to stellar rotation \citep[on both sides of the convective
boundary;][]{Dobler2006,Browning2006}, which has in turn been shown to
depend on both mass and age \citep{Irwin2011}. These magnetic fields
result in chromospheric heating, often observed through the presence
and strength of H$\alpha$ emission \citep[e.g.,][]{Hawley1996}. 

The presence of H$\alpha$ emission has been observed to depend on both
mass (or its proxy, spectral type) and age: early-M dwarfs are active
for a much shorter time ($\sim$1~Gyr) than the mid- ($\sim$4--6~Gyr)
and late-M dwarfs \citep[$>$7~Gyr;][]{West:2008lr}. 
Chromospheric activity strength (usually quantified by the ratio of
the luminosity of H$\alpha$ to the bolometric luminosity, {\lha}) is
also related both to spectral type and age. {\lha} is relatively
constant for M0--M4 dwarfs, then declines with later spectral type
through M9 \citep{West2004,West:2011fk}. The activity trends exhibit significant
scatter compared to their dynamic range and measurement errors. Since
the West et al. studies were based on an ensemble of single-epoch
measurements, the observed spread in activity could originate from the
intrinsic variability of the H$\alpha$ emission line, or could reflect some intrinsic range in magnetic activity strength. 

Repeat observations of H$\alpha$ emission in M dwarfs often show
variability \citep[e.g.,][]{Bopp1978}, but it has only recently been
investigated over both long and short timescales. On timescales of
hours to months, $\sim$80\% of active M dwarfs show significant
variability \citep{Lee2010,Kruse2010}, usually varying by factors of
1.25--4 in H$\alpha$ EW. In data averaged over 150 day windows,
\citet{Gomes-da-Silva2011} found variability on 5--10 year timescales
in 10 out of 30 M0--M5 dwarfs. These variability timescales could be
linked to stellar rotation, the formation and dissipation of active
regions, or long-term variations of the magnetic field. The
relationships between H$\alpha$ variability, stellar mass, and age
have not been thoroughly investigated. \citet{Bell:2012lr}, for
example, found no net trend in H$\alpha$ variability as a function of
spectral type, and an increase in variability with declining activity
strength.

It is unknown how the broad trends in activity strength with respect to mass and age translate to individual systems, and whether the scatter in the observed trends is due to H$\alpha$ variability or is intrinsic to the M dwarf magnetic field strengths. To probe M dwarf activity strength and variability while controlling for mass and age, we have chosen to monitor a sample of near-equal mass, wide M dwarf binary systems selected from the Sloan Low-mass Wide Pairs of Kinematically Equivalent Stars catalog \citep[{\slowpokes};][]{Dhital:2010fk}. With presumably the same age
and metallicity and near-equal masses, these are nearly identical twins. In addition, these are wide 
($\sim$600--15000~AU) binary systems in which both stellar components
have evolved independently, without affecting each other's activity
\citep[as observed in tighter systems, e.g.][]{Meibom2007,Morgan2012}. 

In this paper, we seek to
characterize the differences in activity and variability of {\ha}
emission of M dwarfs by examining {\ha} activity over timescales of
hours to days for these coeval twins. These binary M dwarfs are a unique test case for the relationships between mass, age, and activity. In Section~\ref{Sec:observations}, we describe the observations to
select and characterize our active M dwarf sample in addition to our
time-series spectroscopic observations of the active {\slowpokes}
binaries. We describe the the individual {\ha}
light curves and variability trends in Section~\ref{sec:lightcurve}. In
Section~\ref{sec:AB}, we discuss our H$\alpha$ observations compared to the
mean trends for large samples of M dwarfs. 

\section{Observations and Data Reduction}
\label{Sec:observations}

The {\slowpokes} catalog, constructed from the Sloan Digital Sky
Survey \citep[SDSS;][]{York2000}, comprises 1342 wide ($\geq$500 AU) 
common proper motion pairs with at least one low-mass (K5 or later) component
\citep{Dhital:2010fk}. 
While these systems are classified as binaries based on common proper motion and distance alone, their probability of chance alignment is low. To be part of the {\slowpokes} catalog, each pair was required to have a false positive probability (calculated for each pair using a Galactic model including both stellar density and kinematics) of less than 5\%. \citet{Dhital2012} obtained follow-up observations of a 111 of these pairs, finding that 87\% showed good agreement in their radial velocities. The remaining 13\% of those pairs were likely to be contaminated with low signal-to-noise observations of spectroscopic binaries (providing a challenge for the accurate measurement of radial velocities). As these common proper motion pairs have a high probability of being physically associated, we assume that they are wide binaries. 

With a large number of binaries that
extend down to $\sim$M6 spectral types, the {\slowpokes} catalog is an
ideal source of coeval laboratories to conduct followup studies of M
dwarfs. Because the {\slowpokes} binaries were identified from 
photometry and proper motions without spectroscopic information,
the catalog does not include the magnetic activity as measured by
H$\alpha$ emission for the component stars. Therefore, we selected an
initial sample of {\slowpokes} binaries for spectroscopic
observations. The components of the binaries were restricted to M0 and later dwarfs
($r-z\gtrsim 0.9$) and to have similar $r-z$ colors 
($\Delta(r-z)\leq0.25$), meaning the components are roughly within one-half spectral type. We
also selected for well-resolved components that fit along the length
of the slit (separations of 
7--120$\arcsec$) and a minimum brightness of $r\lesssim$18.  This
resulted in 176 candidate pairs, 36 of which had been determined to be inactive
by \citet{Dhital2012}, giving an initial sample of 140
near-equal mass, M dwarf binaries. While this sample was chosen to be
near-equal mass, there exists a small difference in their $r-z$
colors. Hereafter, we refer to the component with the bluer $r-z$
color as the primary (A) and the one with the redder $r-z$ color as
the secondary (B).                     

\begin{figure}
\centering
\epsfig{file=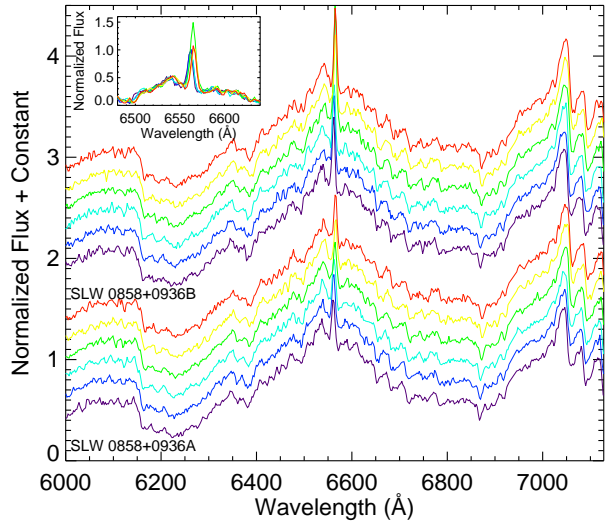,width=\linewidth}
\caption{Red optical spectra (normalized flux as a function of wavelength) for the A and B components of SLW~0858+0936 during the time-resolved observations on UT 2011 Dec 29. Both the A component (bottom six spectra) and the B component (top six spectra) are shown at six different times $\sim$1 hour apart, with time increasing from the bottom (purple) spectrum to the top (red) spectrum. The inset figure shows the H$\alpha$ emission from the B component; the green spectrum was taken during a flare.}  
\label{fig:example}
\end{figure}

Out of the sample of 140 candidate {\slowpokes} binaries, we obtained
optical ($\lambda\sim$5000--10000~{\AA}) spectra for 48 pairs (96
total stars) in our sample using the Dual Imaging Spectrograph (DIS)
on the Astrophysical Research Consortium (ARC) 3.5-m telescope at
Apache Point Observatory (APO). No selection criteria were initially 
applied, but brighter and redder systems were
preferentially observed. The complete sample of observed  M dwarf
binaries is listed in Table~\ref{tab:single_epoch}. DIS uses a dichroic 
to split light into a red and a blue channel; for this study we only
used the red channel. These data were taken with  a 120$\arcsec$ long, 1.5$\arcsec$
wide slit, dispersed with the R300 (2.31~{\AA}/pix) grating, resulting in
an average resolution of R$\sim$1000. Spectra for the components of a
binary were obtained simultaneously by placing both targets on the
spectroscopic slit. Figure~\ref{fig:example} shows spectra of SLW~0858+0936 during time-resolved observations; those spectra are representative of typical spectra obtained during the observations. For each binary, we initially obtained 3 exposures
of 1--10~min each to achieve a minimum total S/N of 20. Flats, biases, and
HeNeAr comparison arcs were also taken at the beginning or end of
every night. The data were bias subtracted, flat fielded, extracted, and wavelength 
calibrated (onto an air scale) using standard IRAF\footnote{IRAF is
  distributed by the National Optical Astronomy Observatories, which
  are operated by the Association of Universities for Research in
  Astronomy, Inc., under cooperative agreement with the National
  Science Foundation} routines. 
 
To identify a sample of {\ha}-active M dwarf binaries for time-resolved
observations, we first determined the spectral types by comparing each
spectrum by eye with templates using the Hammer software \citep{West2004,Covey:2007uq}. In five cases, the spectral type of the A component was one subtype later than that of the B component, in contrast with their $r-z$ colors. We did not reassign types for these systems because of the uncertainties associated with spectral types; since our types are good to $\sim$0.5 subtype, these types are still consistent with a slightly higher mass A component. 
We also used the Hammer to measure the {\ha} equivalent width (EW; hereafter
{\ewha}) for each component. The H$\alpha$ line region is
defined as 6557.61--6571.61~\AA, and the continuum regions are
6500--6550~\AA~and 6575--6625~\AA. The Hammer applies a multi-part
criteria to determine the activity of each object \citep[described in
detail by][]{West2004}. When considering spectra with
S/N$>$3 in the continuum region (a condition met by all our spectra),
those with EW~H$\alpha >0.75$~\AA~are considered active (\textit{y} in
Table~\ref{tab:single_epoch}), those with EW~H$\alpha <0.75$~\AA~and an EW
greater than 3 times its uncertainty are considered weakly active (\textit{w} in
Table~\ref{tab:single_epoch}), and those with {\ewha} below 3 times
its uncertainty are considered inactive (\textit{n} in Table~\ref{tab:single_epoch}). While the resolution of these spectra is somewhat lower than those taken by SDSS (R$\sim$1000 compared to R$\sim$2000) our H$\alpha$ EW measurements should be comparable to those performed on SDSS data. 

Nineteen of the 96  stars exhibited H$\alpha$ in emission. Of the 48 binaries
observed, eight exhibited H$\alpha$ emission in both components, and three exhibited H$\alpha$ emission in one component. In
addition to measuring {\ewha}, we also calculated the spectral
type-independent measure of activity strength ({\lha}) for the active
dwarfs. The conversion from {\ewha} to {\lha} relies on the $\chi$
factor, calculated by \citet{West2008a} as a function of $i-z$ color
or spectral type. We adopted $\chi$ as a function of spectral type to calculate {\lha} for each object. Spectral types, EW
H$\alpha$, and {\lha} are given in Table~\ref{tab:single_epoch}.  

We selected the five brightest of the eight binaries with two active components for
time-resolved spectroscopy. These five systems were each observed for
1--5 hrs per night for a total of 47 hours over 11 nights. We used the
same setup for the time-resolved observations as for the single epoch observations, with typical
exposure times of 1--5 minutes for each system. The UT dates and hours
observed are given in Table~\ref{tab:obs}. Four of the systems were
observed on three or four different nights while SLW~2315$-$0045 was
only observed on one night. We measured {\ewha} using the
Hammer for each spectrum; the H$\alpha$ variations are discussed for
each pair in the next section.

\setcounter{table}{1}

\begin{deluxetable}{ccccc}
\tabletypesize{\scriptsize}
\tablecaption{Log for time-domain observations \label{tab:obs}}
\tablewidth{0pt}
\tablehead{
  \colhead{UT} &
  \colhead{Designation} &
  \colhead{Total t$_{\textrm obs}$}&
  \multicolumn{2}{c}{\# flares}  \\
    \colhead{} &
  \colhead{} &
  \colhead{(hrs)}&
  \colhead{A}  &
  \colhead{B}   }
\startdata
2011 Feb 18 &SLW 0741$+$1955AB & 0.31 & 0 & 0 \\
2011 Mar 09 & SLW 0858$+$0936AB & 4.48 & 0 & 1 \\
2011 May 02 & SLW 1120$+$2046AB & 3.80 & 0 & 0 \\
2011 Aug 27 & SLW 2315$-$0045AB & 1.03 & 0 & 0 \\
  	   & SLW 0149$+$2215AB & 0.07 & 0 & 0 \\
2011 Oct 15 & SLW 0149$+$2215AB & 3.40 & 0 & 1 \\
2011 Oct 24 & SLW 0741$+$1955AB & 0.80 & 0 & 0 \\
            & SLW 0149$+$2215AB & 3.83 & 0 & 0 \\
2011 Nov 28 & SLW 0741$+$1936AB & 5.20 & 0 & 1 \\
2011 Dec 29 & SLW 0858$+$0936AB & 5.53 & 0 & 1 \\
2012 Feb 16 & SLW 1120$+$2046AB & 4.73 & 0 & 0 \\
2012 Feb 21 & SLW 1120$+$2046AB & 2.30 & 0 & 0 \\
2012 Mar 14 & SLW 1120$+$2046AB & 6.92 & 0 & 1 \\
2012 Mar 24 & SLW 0858$+$0936AB & 3.34 & 0 & 0 \\
            & SLW 0741$+$1936AB & 1.37 & 0 & 0 
\enddata
\end{deluxetable}

\setlength{\tabcolsep}{2pt}
\begin{deluxetable}{cc cc cccccc}
\tabletypesize{\scriptsize}
\tablewidth{0pt}
\tablecaption{Binaries with time-domain spectra\label{tab:sample}}
\tablehead{
  \colhead{}	&	
  \colhead{}	&	
\multicolumn{2}{c}{Including Flares}&
\multicolumn{2}{c}{Excluding Flares}
\\
  \colhead{Designation}	&	
  \colhead{Comp}	&	
  \colhead{$\langle${\ewha}$\rangle$}  &
  \colhead{$\frac{\sigma_{\ewha}}{\langle{\ewha}\rangle}$} &
  \colhead{$\langle${\ewha}$\rangle$} &
  \colhead{$\frac{\sigma_{\ewha}}{\langle{\ewha}\rangle}$} 
\\
  \colhead{SLW$+$}	&	
  \colhead{}	&	
  \colhead{(\AA)}  &
  \colhead{} &
  \colhead{(\AA)} &
  \colhead{}
}
\startdata
0149$+$2215	& A   & 2.17	$\pm$ 0.30 & 0.14 & 2.17 $\pm$ 0.30 & 0.14\\
               		& B   & 4.56	$\pm$ 1.43 & 0.31 & 4.10 $\pm$ 1.13& 0.27\\
0741$+$1955	& A   &  1.95	$\pm$ 0.57  & 0.30 & 1.47 $\pm$ 0.26 & 0.18\\
             		& B  & 5.25	$\pm$ 0.80 & 0.15 & 4.86 $\pm$ 0.43 & 0.09\\
0858$+$0936	& A  & 2.66	$\pm$ 0.39 & 0.15 & 2.66 $\pm$ 0.39& 0.15\\
             		& B  & 4.90	$\pm$ 2.07  & 0.42 & 4.35 $\pm$ 0.84& 0.19\\
1120$+$2046	& A  & 1.65	$\pm$ 0.74 & 0.45 & 1.65 $\pm$ 0.74 & 0.45\\
          		& B   & 5.49	$\pm$ 0.91  & 0.17 & 5.24 $\pm$ 0.54 & 0.10 \\
2315$-$0045	 & A  & 3.96	$\pm$ 0.31& 0.08 & 3.96 $\pm$ 0.31  & 0.08\\
             		& B   & 4.01	$\pm$ 0.26  & 0.07 & 4.01 $\pm$ 0.26 & 0.07
\enddata	
\end{deluxetable}

\section{Characterizing H$\alpha$ Lightcurves}
\label{sec:lightcurve}
\begin{figure*}
\centering
\epsfig{file=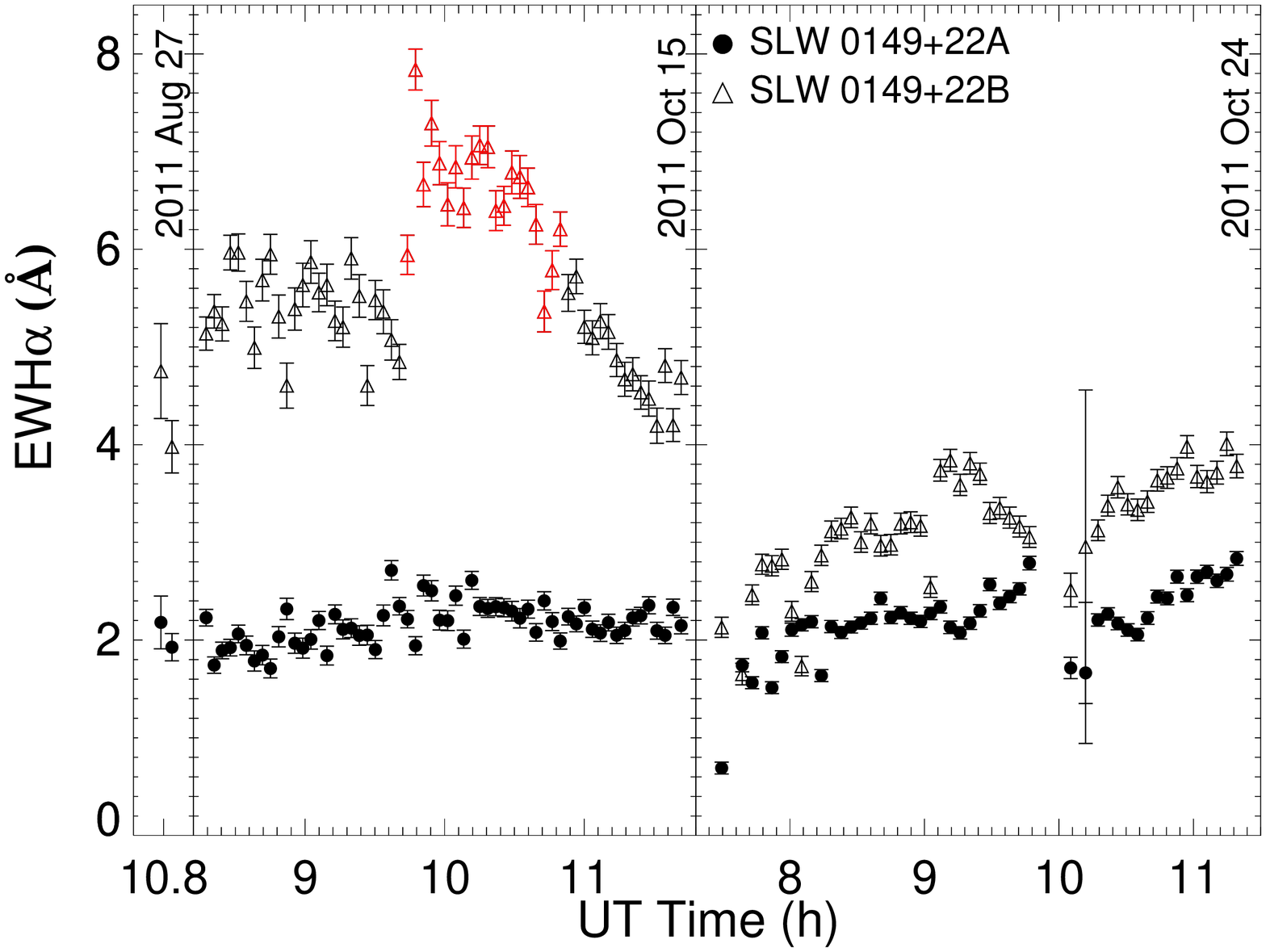,angle=90,width=0.49\linewidth}
\epsfig{file=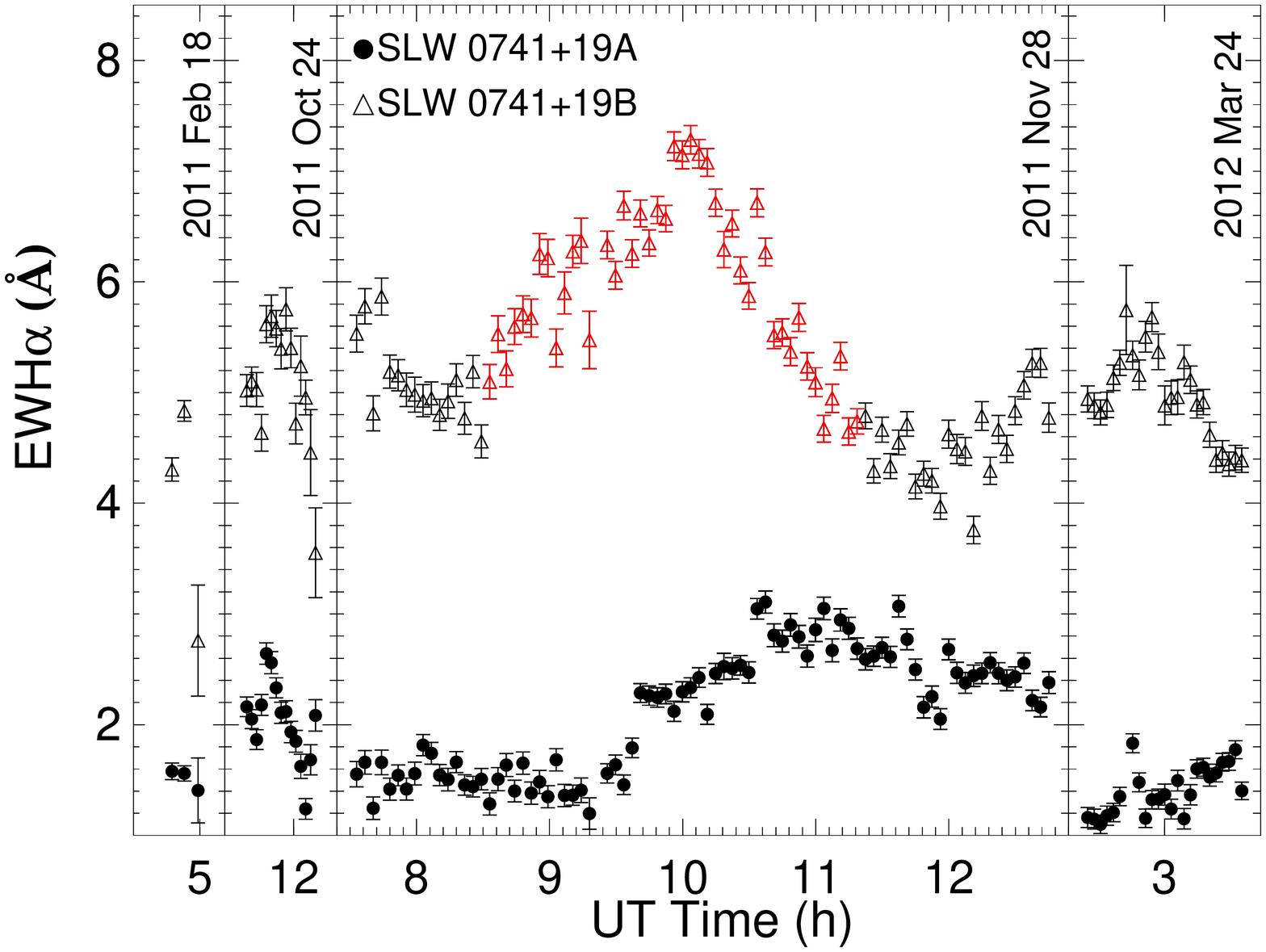,angle=90,width=0.49\linewidth}
\epsfig{file=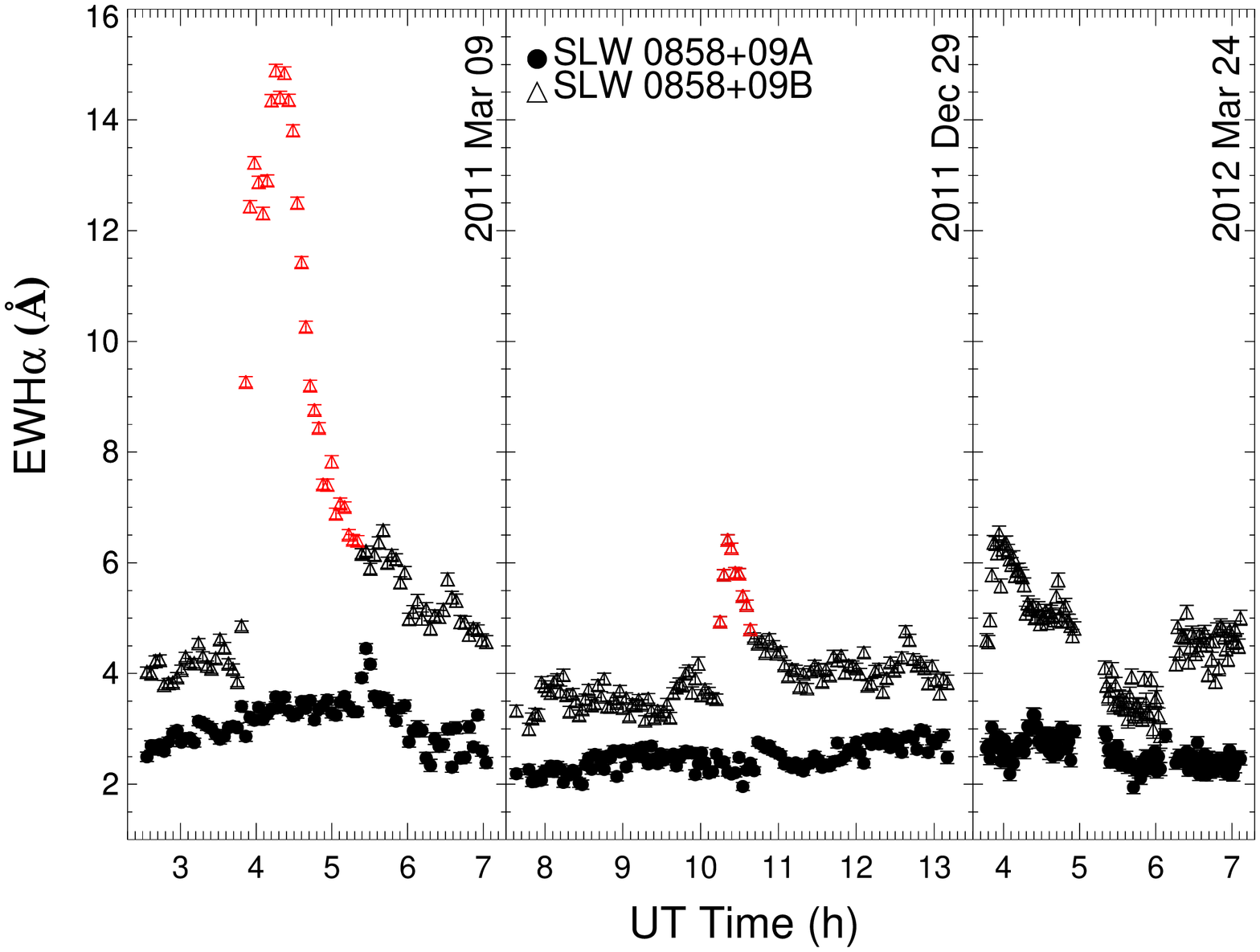,angle=90,width=0.49\linewidth}
\epsfig{file=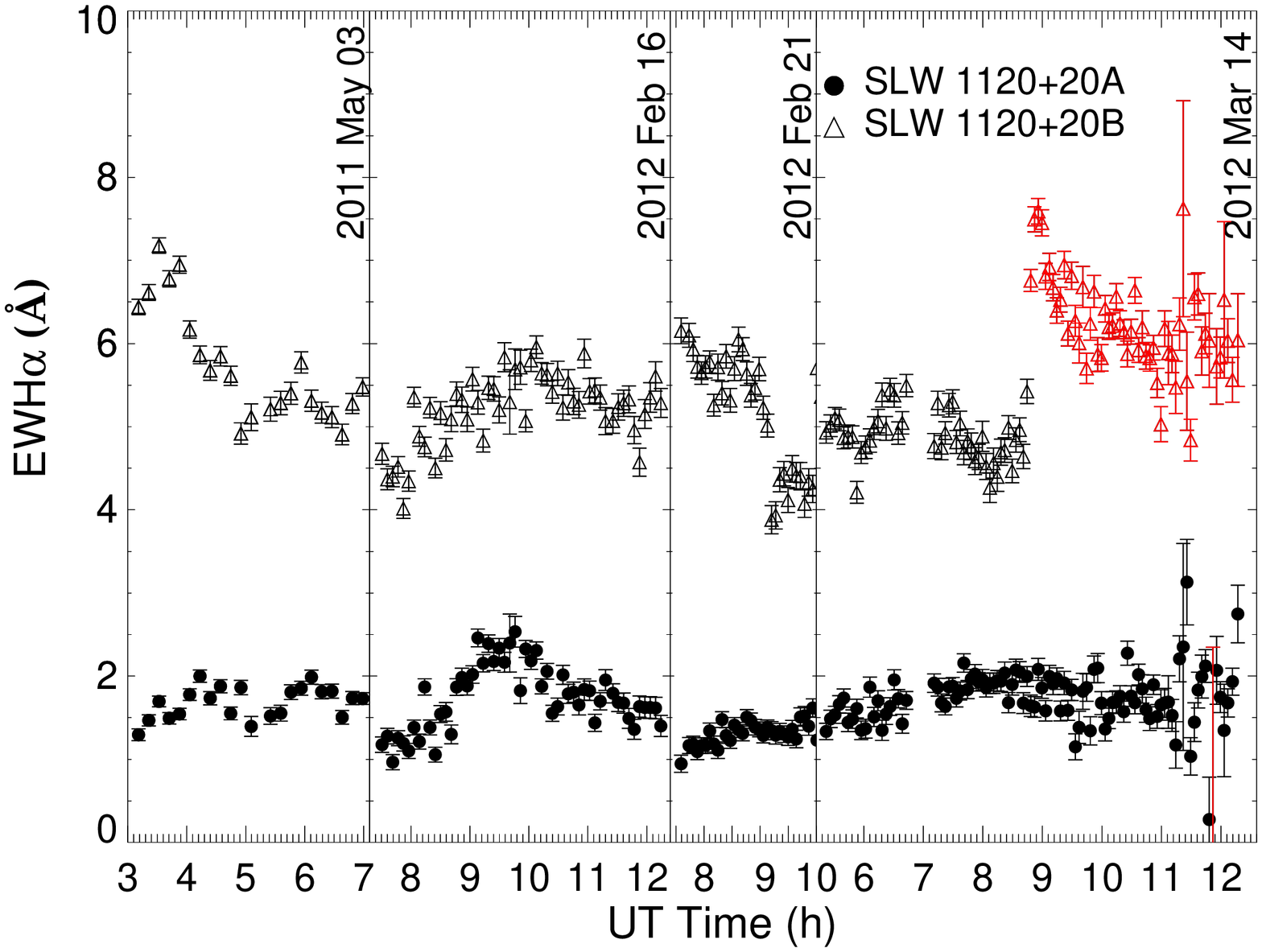,angle=90,width=0.49\linewidth}
\epsfig{file=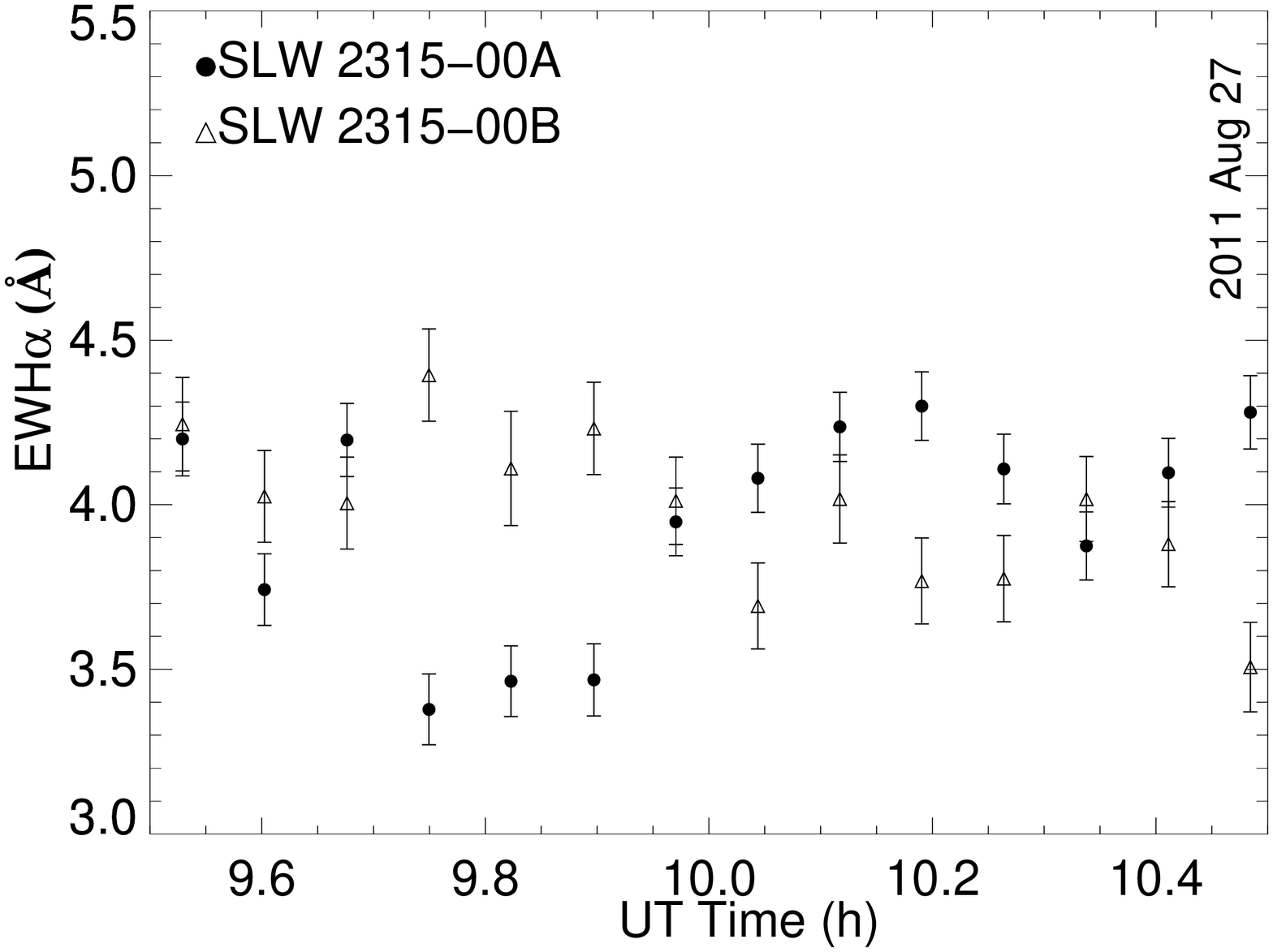,angle=90,width=0.49\linewidth}
\caption{{\ewha} as a function of time for each of the binaries.
 (upper left): SLW 0149+22, (upper right): SLW 0741+19,
 (middle left): SLW 0858+09, (middle right):  SLW
  1120+20, (bottom): SLW 2315-00. Each panel is labeled with
  the UT dates when the binary was observed. The B component data are shown as open
circles and the A component data are filled circles. The red data indicate
  the portion of the curve that has been classified as a flare event,
  as described in Section~\ref{sec:flare}. Formal uncertainties are
  shown on each point; many are approximately the size of the point. These stars have separations far greater than those of interacting binaries, so any variations that are similar between the two stars are either coincidental or due to observational effects (e.g., changing airmass). 
}   
\label{fig:HA}
\end{figure*}

Our initial goal was to
determine if observed differences in {\ewha} between the two stars in each binary
were due to variability, e.g., two objects with the same mean
{\ewha} would appear different in single epoch observations if their instantaneous measured values differed from the mean. The light curves from the
time-resolved observations are shown in Figure~\ref{fig:HA}. Our
initial examination of the light curves revealed the presence of
flares, so we identified and removed the flares before examining
the quiescent light curves in more detail.

\subsection{Removing Flares}
\label{sec:flare}
Both flares and quiescent activity are caused by the interaction of
strong surface magnetic fields with the stellar atmosphere, but the
details of that interaction are likely to be different. Flares are
thought to be triggered by magnetic reconnection events resulting in
dramatic heating over short timescales \citep[e.g.,][]{Cram1979},
while quiescent activity is characterized by emission over longer
timescales from lower temperature material
\citep[e.g.,][]{Robinson1990}. Our primary focus is the quiescent
variability, but H$\alpha$ emission traces both quiescent emission and
flares. We calculated the mean ($\langle{\ewha}\rangle$) and standard deviation ($\sigma_{\rm EW}$) of the H$\alpha$
emission both from the
entire light curves and from the light curves with flares
removed. While flares have been traditionally identified ``by eye"
\citep[e.g.,][]{Pettersen1984}, to clearly separate flares from
smaller-scale variations we chose to remove flares using the
quantitative method developed by \cite{Hilton:2011th}, which we summarize here.  

\begin{figure}[h]
\centering
\epsfig{file=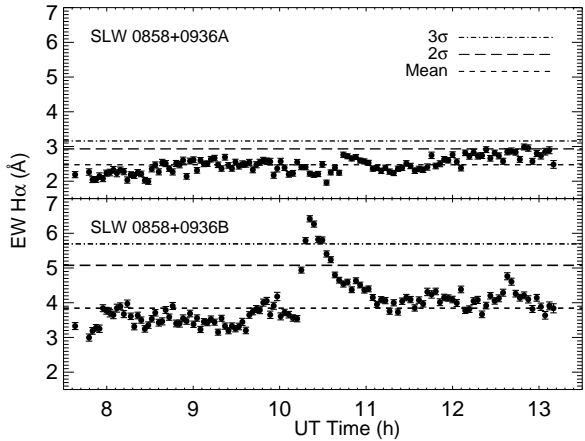,angle=90,width=\linewidth}
\caption{{\ewha} as a function of time for SLW 0858+0936A (top) and SLW 0858+0936B (bottom) on UT 2011
Dec 29. The data (black filled circles with error bars) are shown along with
  the mean and multiples of the standard deviation (dashed
  lines). According to the flare criteria described in
  Section~\ref{sec:flare}, the A component shows only quiescent
  variations while the B component flares at $t = 10.3$~hr. } 
\label{fig:flares}
\end{figure}

We first determined a quiescent mean and standard deviation {\ewha} for each binary component on 
each night of data. For stars that appeared to exhibit a flaring event, we used 
the quiescent period before and/or after the flare to obtain our $\langle{\ewha}\rangle$ and $\sigma_{\rm EW}$. We then defined a flare as an event with at least one
point above 3$\sigma$ and at least five points above 2$\sigma$ of the
quiescent $\langle{\ewha}\rangle$. Figure~\ref{fig:flares} shows a subset of the {\ewha} measurements of
 SLW~0858$+$0936AB (from UT 2011 Dec 29) with the
quiescent mean as well as the 1$\sigma$ and 2$\sigma$ deviations given
as dashed lines. The A component shows variations, but none are large
enough to meet the flare criteria. The B component also varies
throughout the night, with one of the variations being large enough
to be considered a flare.

Using these criteria, we initially identified six flares. We then
confirmed each flare by eye, accepting five flares and rejecting one
spurious flare (on SLW 0741+19B; see discussion below), resulting in a total of five flares out of 47 hours of
monitoring of the five binary systems (comparable to 94 hours on
single M dwarfs). All five of the flares occurred on the B components
of the binaries, including two on SLW~0858+09B. The number of flares
on each star on each night is noted in Table~\ref{tab:obs} and the
flares are  indicated in the lightcurves shown in Figure~\ref{fig:HA}.

\subsection{EW H$\alpha$ Light Curves}
The EW H$\alpha$ light curves shown in Figure~\ref{fig:HA} are described in detail below. The $\langle{\ewha}\rangle$ and $\sigma_{\rm EW}$ were calculated both with and without flares and are given in Table~\ref{tab:sample}. Here we discuss the properties with flares removed. 

{\bf SLW 0149+2215:} A total of 7.3 hours of data were obtained for this system over
the course of three nights. The A component maintained a relatively
constant {\ewha} of $\langle$EW H$\alpha \rangle=2.17$~\AA~with $\sigma_{\rm EW} = 0.30$~\AA. H$\alpha$ emission from the B component was more variable,
showing a 2~\AA~variation between the two longer nights of
observations with $\langle$EW H$\alpha \rangle=4.56$~\AA~and
$\sigma_{\rm EW}=1.43$~\AA. It is unclear whether the base value
surrounding the flare on UT 2011 Oct 15 is true quiescence or part of a longer trend of
elevated activity, but it is clear that the B component has stronger
and more variable H$\alpha$ emission than the A component. 

{\bf SLW 0741+1955:} We observed this system for 7.7 total hours on
four different nights. The emission from A component was relatively constant with an $\langle$EW H$\alpha \rangle=1.95$~\AA~and $\sigma_{\rm EW}=0.57$~\AA. The emission from the B component was both stronger ($\langle$EW H$\alpha \rangle=5.25$~\AA) and more variable ($\sigma_{\rm EW}=0.80$~\AA) than the A component. 
The emission from the B component included a flare, while
the emission from the A component did not. At $t=9.8$~hr, the
A component showed an increase that was initially marked as a flare, but it 
was rejected when reviewed by eye because it lacked a decay 
phase after the initial rise.

{\bf SLW 0858+0936:} This system was observed for 13.4 hours on three separate nights. The H$\alpha$ emission from the A component
was relatively strong ($\langle$EW H$\alpha \rangle = 2.66$~\AA) but not variable
($\sigma_{\rm EW}=0.39$~\AA). The H$\alpha$ emission from the B
component was very strong and variable ($\langle$EW$\rangle$ = 4.90\AA;
$\sigma_{\rm EW}=2.07$~\AA) even with the two flares excluded. On UT 2011 Mar 9, the H$\alpha$ 
emission showed a large flare that peaked at EW H$\alpha=15$~\AA~and on UT
2011 Dec 29, a small flare peaked at an EW of H$\alpha=6$~\AA. 

{\bf SLW 1120+2046:} This binary was observed for a total of 17.8 hours
during four nights. The H$\alpha$
emission from the A component was the weakest of all the observed M
dwarfs ($\langle$EW H$\alpha \rangle = 1.65$~\AA) but more variable than other A components
($\sigma_{\rm EW}=0.74$~\AA). The mean H$\alpha$ emission from the B
component was stronger than that of other B components ($\langle$EW H$\alpha \rangle = 5.24$~\AA) with relatively low variability ($\sigma_{\rm EW}=0.54$~\AA; similar to SLW 0741+19B). On the last
night of observations for this pair, we observed a small flare on the B component which
peaked at EW H$\alpha \sim 8$~\AA.  

{\bf SLW 2315$-$0045:}  This binary pair was both the latest-type dM in
our sample (M5/M5) and has the least observations (one hour on a single
night).  This binary had the smallest difference in strength
and variability between the two components, with values of
$\langle$EW H$\alpha \rangle \sim4$~\AA~and $\sigma_{\rm EW}\sim0.3$~\AA~for
both A and B. During the short duration of out observations, SLW 2310$-$0045 was less variable than the other observed systems. 

\subsection{Variability}
\begin{figure*}
\centering
\epsfig{file=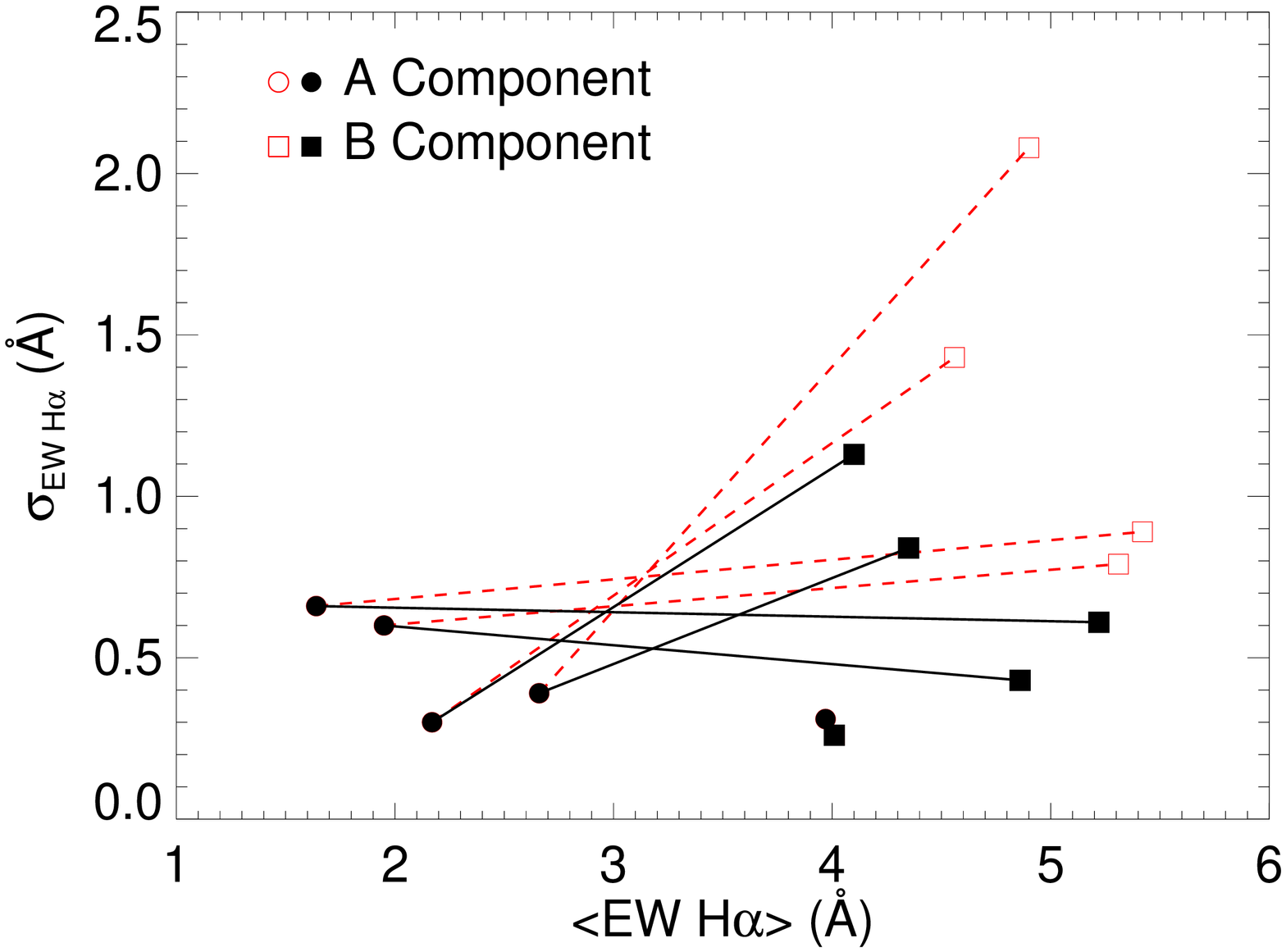,angle=90,width=0.49\linewidth}
\epsfig{file=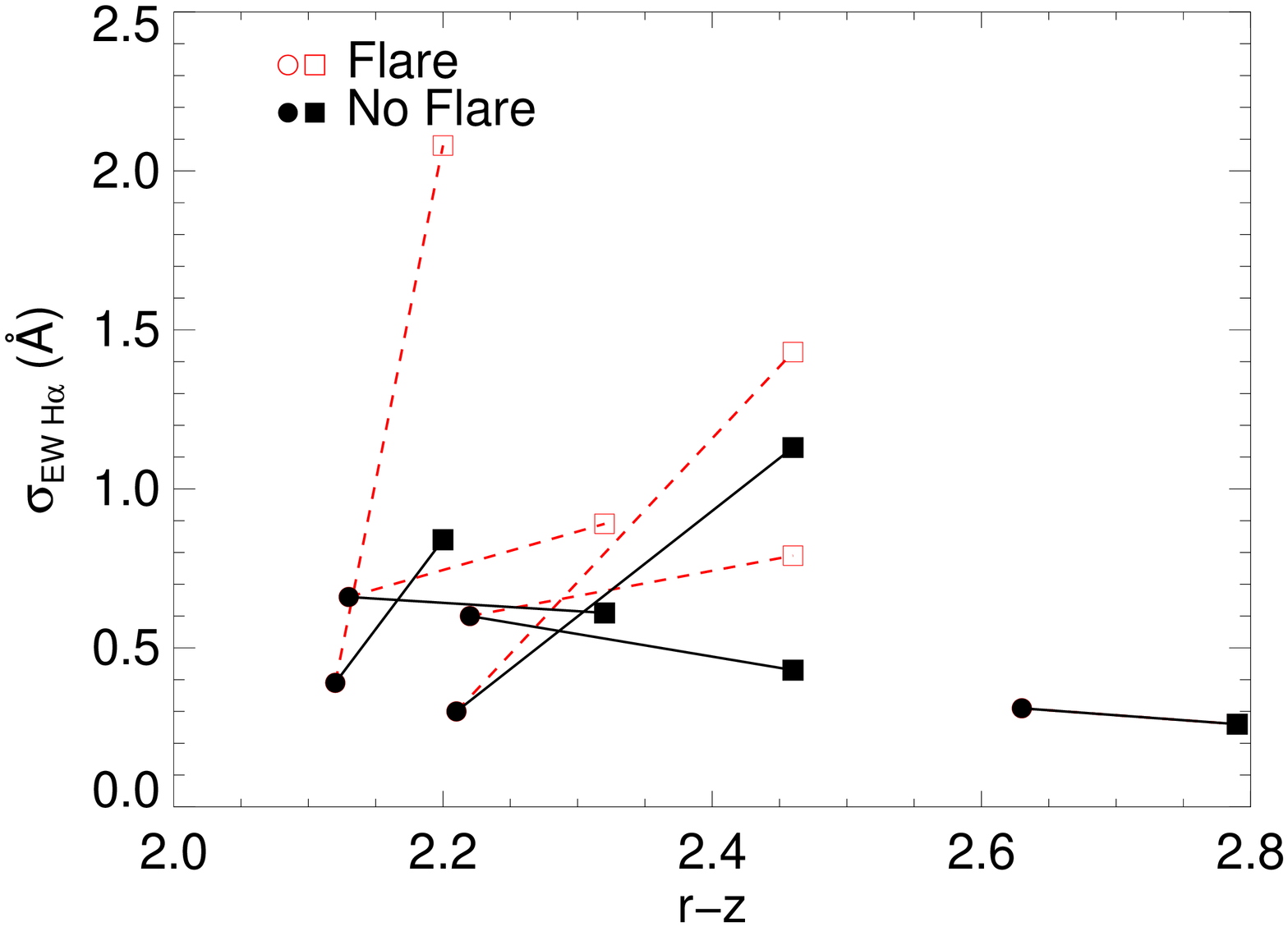,angle=90,width=0.49\linewidth}
\epsfig{file=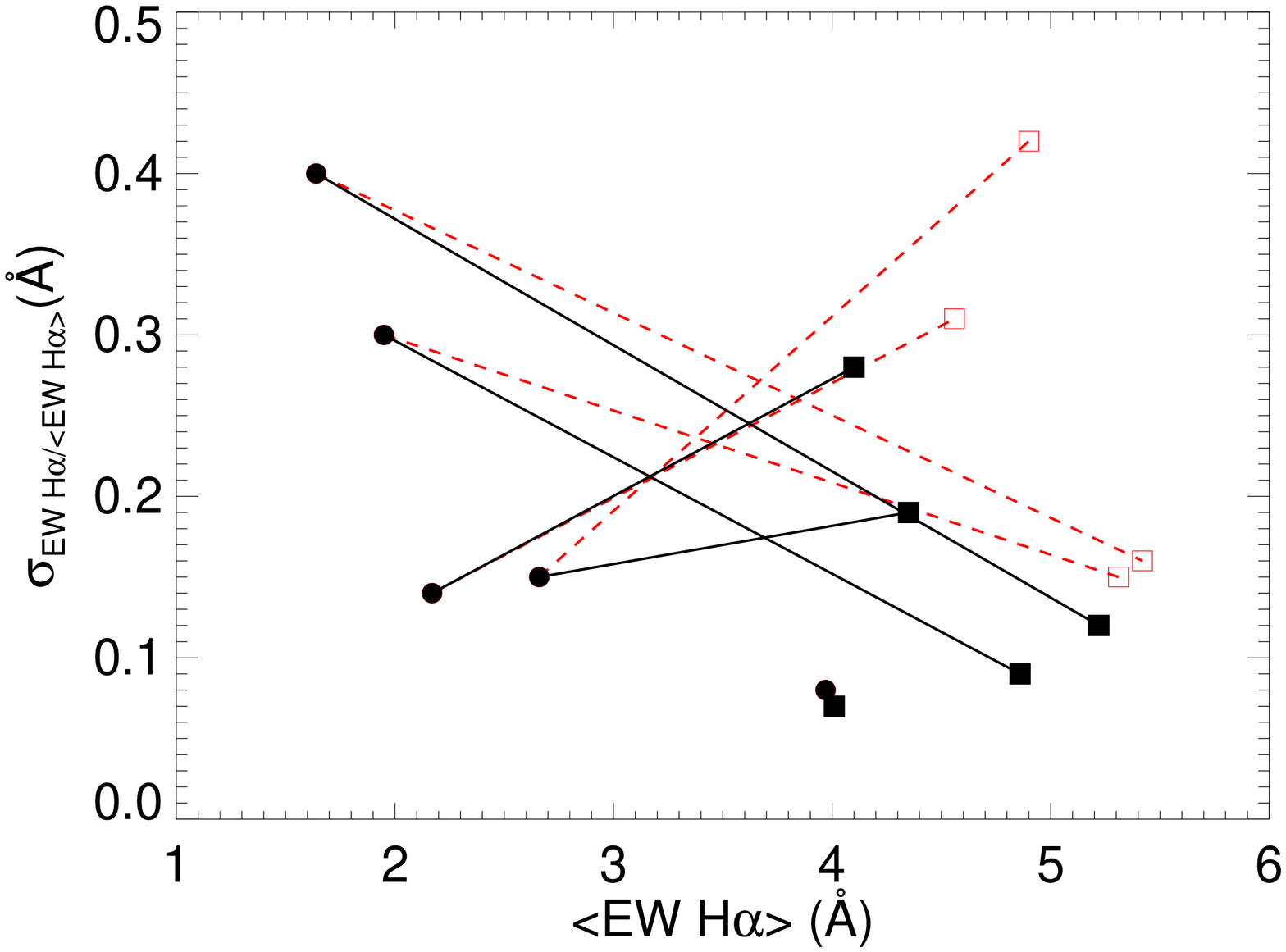,angle=90,width=0.49\linewidth}
\epsfig{file=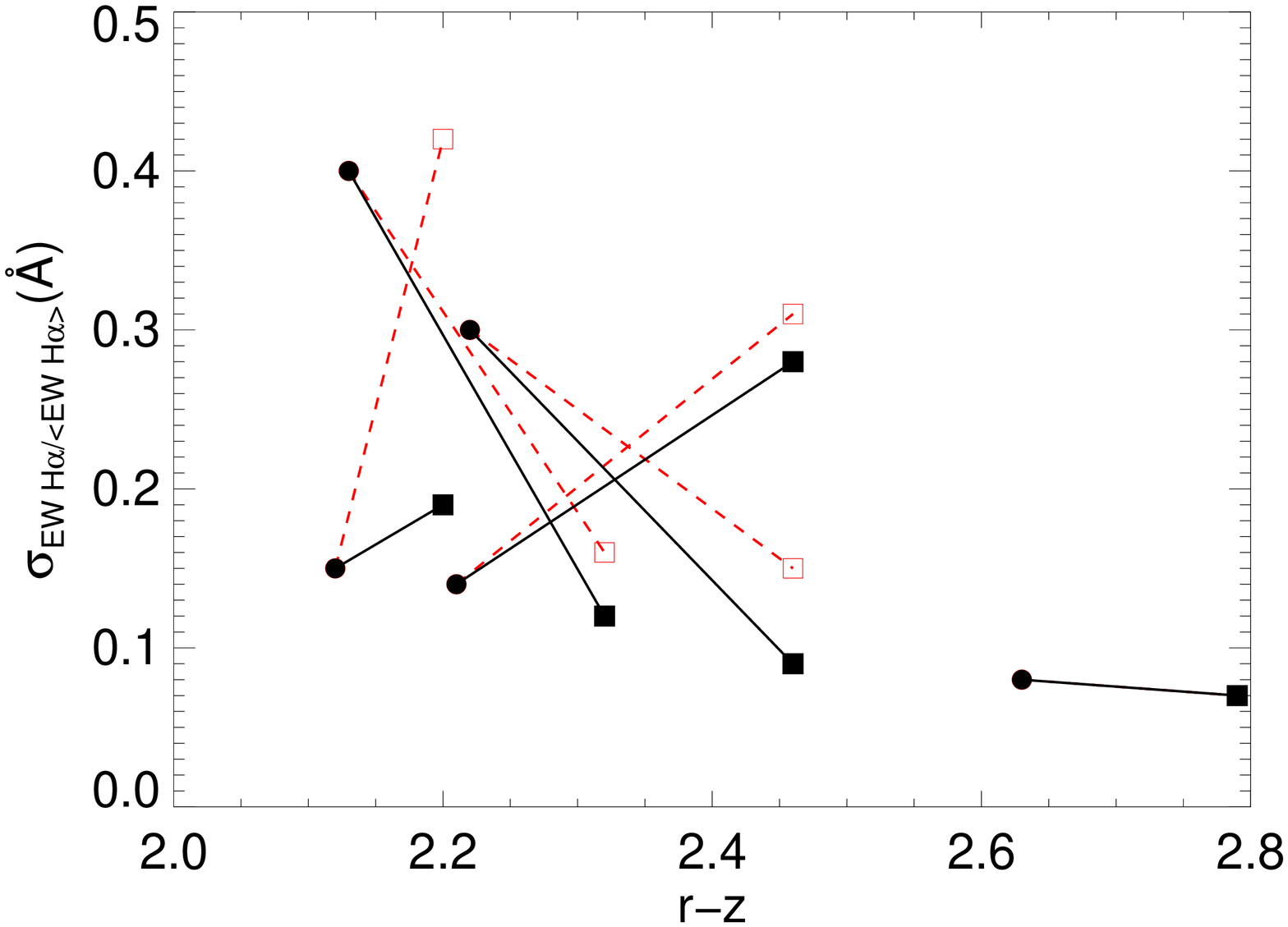,angle=90,width=0.49\linewidth}
\caption{Variability of the twins with time resolved observations. The
  top two panels show the variability of {\ewha} ($\sigma_{\ewha}$) as
  a function of the mean {\ewha} ($\langle{\ewha}\rangle$; top left
  panel) and $r-z$ color (top right panel). The bottom two panels show
  the normalized variability of {\ewha}
  ($\sigma_{\ewha}/\langle{\ewha}\rangle$) as a function of the mean
  {\ewha} ($\langle{\ewha}\rangle$; bottom left panel) and $r-z$ color
  (bottom right panel). Each A (circles) and B (squares) component
  is connected, and values computed without flares (solid black
  symbols) are distinguished from values computed with flares (open
  red symbols).} 
\label{fig:var}
\end{figure*}

The production of H$\alpha$ emission by an active chromosphere is a dynamic process on timescales of minutes to decades. Several groups have used different statistics to characterize that variability; the ratio of maximum to minimum {\ewha} \citep{Kruse2010,Lee2010}, the standard deviation of {\ewha} \citep{Gizis:2002qy} and normalized standard deviation of {\ewha} \citep{Bell:2012lr}.We choose characterize variability with the standard deviation of {\ewha} ($\sigma_{\ewha}$) and normalized variability with $\sigma_{\ewha}$ divided by $\langle{\ewha}\rangle$. Table~\ref{tab:sample} includes these quantities for each of the twins with time-resolved data, and they are shown with respect to mean {\ewha} and $r-z$ color in Figure~\ref{fig:var}. The $\sigma_{\ewha}$ and $\sigma_{\ewha}/\langle{\ewha}\rangle$ are slightly higher than the mean values reported by \citet{Bell:2012lr} for M dwarfs with similar spectral types and emission levels, which could be due to the differences in cadence and timescale of the observations. 

The un-normalized variability is on average higher for the B component than the A component. This is, perhaps, simply due to the B components having stronger activity than their twins. More H$\alpha$ emission allows a larger dynamic range for variation. Increasing variability with larger $\langle{\ewha}\rangle$ is not visible when the normalized variability is instead examined. The four pairs with strong variability are evenly divided between the A and B components showing stronger variability. If variability was a strong function of age, we would expect the co-eval companions to have similar variability, but our sample reveals no relationship between age, activity level, and variability. If there are correlations between these quantities, larger samples of M dwarfs or longer timescale H$\alpha$ monitoring may be needed to detect them.  

\section{M dwarfs Activity Trends With Mass and Age}
\label{sec:AB}
\begin{figure}
\centering
\epsfig{file=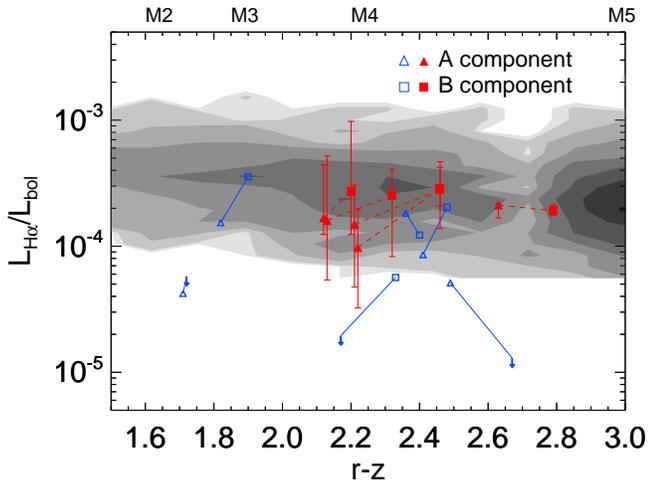,angle=90,width=\linewidth}
\caption{The ratio of the H$\alpha$ luminosity to the bolometric
  luminosity as a function of $r-z$ color for the active binaries in
  our sample. The distribution of active SDSS M dwarfs is also shown
  for comparison  \citep[contours;][]{Bell:2012lr}. A (triangle) and B
  (square) components of each pair are connected (solid and
 dashed lines). For stars with single epoch data (blue), the
  measured $L_{\rm H\alpha}/L_{\rm bol}$ is shown while for pairs with
  time-resolved data (red), both the median and total spread in
  $L_{\rm H\alpha}/L_{\rm bol}$ are shown. For the three pairs
  consisting of one active and one inactive component, the upper limit
  of H$\alpha$ emission (the measured EW from the highest individual
  spectrum, which was not high enough to meet our criteria for active)
  is shown as a down arrow.}
\label{fig:lhlb_rz}
\end{figure}

Comparing the H$\alpha$ emission of the A and B components to overall trends can provide unique constraints on the relationship between M dwarf mass, age, and activity. Figure~\ref{fig:lhlb_rz} shows the activity strength ($L_{\rm H\alpha}/L_{\rm bol}$) of the active M dwarfs from \cite{West:2011fk} as a function of $r-z$ color \citep[often a proxy for spectral type;][]{Bochanski2011}. As expected, the activity strength declines slowly with decreasing mass (redder $r-z$ color) for these mid-M dwarfs \citep{Hawley1996,Gizis:2002qy,West:2008lr,West:2011fk}, though with relatively large scatter: between $1.5<r-z<3.0$, there is at least an order of magnitude range of activity strengths ($L_{\rm H\alpha}/L_{\rm bol}\sim0.0001$ to 0.001). We also show the activity strength of the eleven binaries with at least one active component (including upper limits for the inactive component in those pairs). Of the eleven binaries, four have a more active A component, while the rest have more active B components. The binaries pairs do not, on average, follow the ensemble trend of decreasing activity strength with redder $r-z$ color. 

\begin{figure}
\centering
\epsfig{file=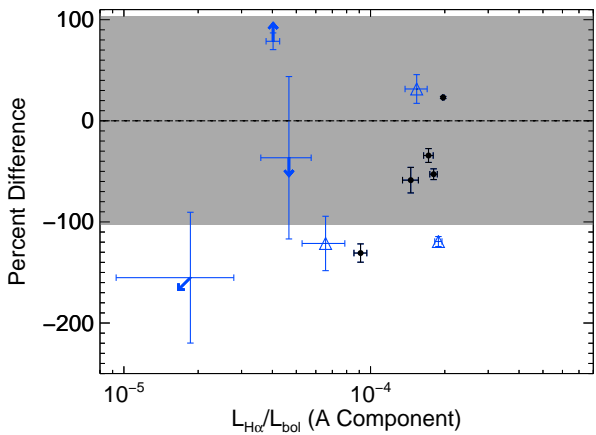,angle=90,width=\linewidth}
\caption{The percent difference in activity strength between A and B
  components for each pair as a function of the A component activity
  strength. Here, systems with a positive percent difference have
  stronger activity in the A component. Binaries with
  time-resolved observations (black circles) are distinguished from
  those with only a single set of observations (blue triangles). Systems with only one active component are represented with arrows denoting upper or lower limits; for SLW1446+5324, the limit is on both the A component H$\alpha$ and the difference between the A and B component. For comparison, the normalized standard deviation of the ensemble activity
  level for the M0-M5 dwarfs from \citet{Bell:2012lr} is shown (grey shaded region). Systems falling outside the grey shaded region have significantly different levels of activity compared to the typical range of the ensemble. 
} 
\label{fig:sameness}
\end{figure}

To understand the significance of the differences between the A and B components compared to the scatter in emission strength for the ensemble of measurements, in Figure~\ref{fig:sameness} we show the percent difference between the A and B components' {\lha} as a function of the A component's {\lha} compared to the normalized standard deviation of \lha\ values from \citet{Bell:2012lr}. If magnetic activity monotonically decayed for both A and B twin stars together, we would naively expect all active pairs to have a difference close to zero on the vertical axis. However, this zero difference only falls within the uncertainties of one system. This is true even of the five pairs with dedicated monitoring; the variability over timescales of hours to months does not bring the A and B emission strengths closer together. While no clear trend is seen as a function of the A component's \lha, the difference in activity spans from $-$160 to 80\%, with the majority of systems having stronger H$\alpha$ emission in the slightly redder B component.

Additionally, four of the pairs in our sample show differences in their activity significantly greater than the typical scatter from \citet{Bell:2012lr}, all in which the B component has stronger activity. These are particularly interesting systems that would benefit from additional observations. It is possible that despite their coevality, these twin M dwarfs have different rotation rates and magnetic field strengths. Another possibility is the existence of Solar-like activity cycles that could be revealed though additional H$\alpha$ monitoring. 

While there is not evidence for a strong trend between activity and age, our data are consistent with the ``activity lifetime" model of M dwarf H$\alpha$ emission. Our full sample included 37 inactive pairs, three pairs with one active component and eight pairs with two active components.  The 37 systems with two inactive components may be older than the active lifetimes for their spectral types. Based on the \citet{West:2008lr} age-activity relations, this places lower limits on the stellar ages of $\sim$1 Gyr for M0--M2 dwarfs, and 2--7 Gyr for M3--M5 dwarfs, which are typical of disk stars. 

In the three binary systems from our sample for which only one stellar component showed H$\alpha$ activity, the active components (two A components and one B component) had relatively weak H$\alpha$ emission (EW $<0.5$\AA).  Given the moderate S/N and low spectral resolution of our observations, very weak emission from the ``inactive'' components may have been below our measurement sensitivity. These systems may be in fact be undergoing the transition from active to inactive states. This would in turn imply that the binary's age is approximately at the expected activity lifetime (e.g., $\sim$4 Gyr for M4). The relative scarcity of these systems in our dataset indicates that this transition must occur relatively quickly.

Variability may instead be a cause of the scatter in activity strength with respect to mass and age. If variability is the main cause for the scatter, we would expect the activity of all mid-M dwarfs to be about equal when observed over long timescales. For the binaries with time-resolved data, we plot the median value and the full range of observed activity strengths in Figure~\ref{fig:sameness}. These ranges show the possible range of values that would be observed for each M dwarf in a single spectrum. This variability ranges from a factor of two to an order of magnitude and is comparable to the full range of scatter in the singe-epoch data, indicating that variability could be the main cause for the scatter.  Variability does not, however, explain the differences between the A and B components because they persist on timescales of weeks to years. To understand the different H$\alpha$ strengths of coeval binaries, additional observations are needed. 

\section{Summary}
\label{sec:sum}
We presented optical spectra for 96 early- to mid-M dwarfs found in 48 coeval twin binary systems. Of those binaries, eight had H$\alpha$ emission from both components and three showed H$\alpha$ emission from one component. The active wide binary components did not show the similar levels of H$\alpha$ emission that would initially be expected from M dwarfs of the same mass and age. Instead, the A and B components exhibited different levels of emission, with four of the eleven having differences larger than the standard deviation of the scatter in $L_{H\alpha}/L_{bol}$ for the \citet{West:2011fk} M dwarfs. Despite the mean trend of decreasing activity strength with decreasing mass, we found that seven of the eleven binaries had stronger activity on the less massive (B) component. 

We also presented 47 hours spectroscopic monitoring for five of the wide binaries. We examined their H$\alpha$ light curves in detail, identifying five flares that all occurred on the B components of the binaries. The majority of these M dwarfs exhibited significant variability; the range of activity strength was comparable to the scatter in {\lha} as a function of $r-z$ color. Over the $\sim$1 year timescale of the time-resolved observations, the observed binaries continued exhibit stronger activity from their B components, but variations on longer timescales (e.g., decades) may further modify their relative activity strengths. The variability from our time-resolved data showed no trends with either $r-z$ color or H$\alpha$ EW; we find that variability is independent of both stellar mass and activity strength. 

While M dwarf ensemble trends indicate strong relationships between mass, age, and the presence and strength of activity, our observations reveal a complicated relationship between mass, age, and activity for individual stars. The differences in activity in coeval twins could indicate long timescale Solar-type variations in the magnetic field strength or could be due to intrinsic differences in rotation rate and magnetic field generation. Time-resolved observations over longer timescales, in addition to more detailed studies of coeval twins, will be instrumental in a complete understanding of M dwarf activity.

\acknowledgements
J.\ R.\ A.\ D.\ acknowledges funding from NASA ADP grant NNX09AC77G. 
S.\ D.\ acknowledges funding from NSF grant AST-0909463. 
A.\ A.\ W.\  acknowledges funding from NSF grants AST-1109273 and
AST-1255568 and also the support of the Research Corporation for
Science Advancement's Cottrell Scholarship.

Our work relies on data from the Sloan Digital Sky Survey. Funding for
the SDSS and SDSS-II has been provided by the Alfred P. Sloan
Foundation, the Participating Institutions, the National Science
Foundation, the U.S. Department of Energy, the National Aeronautics
and Space Administration, the Japanese Monbukagakusho, the Max Planck
Society, and the Higher Education Funding Council for England. The
SDSS Web Site is http://www.sdss.org/. 

The SDSS is managed by the Astrophysical Research Consortium for the
Participating Institutions. The Participating Institutions are the
American Museum of Natural History, Astrophysical Institute Potsdam,
University of Basel, University of Cambridge, Case Western Reserve
University, University of Chicago, Drexel University, Fermilab, the
Institute for Advanced Study, the Japan Participation Group, Johns
Hopkins University, the Joint Institute for Nuclear Astrophysics, the
Kavli Institute for Particle Astrophysics and Cosmology, the Korean
Scientist Group, the Chinese Academy of Sciences (LAMOST), Los Alamos
National Laboratory, the Max-Planck-Institute for Astronomy (MPIA),
the Max-Planck-Institute for Astrophysics (MPA), New Mexico State
University, Ohio State University, University of Pittsburgh,
University of Portsmouth, Princeton University, the United States
Naval Observatory, and the University of Washington.

\setcounter{table}{0}

\LongTables

\setlength{\tabcolsep}{3pt}
\begin{deluxetable*}{ccccrrcrcccc}
\tabletypesize{\scriptsize}
\tablewidth{0pt}
\tablecaption{List of all observed binaries\label{tab:single_epoch}}
\tablehead{
  \colhead{Designation} &
  \colhead{Comp}	&	
  \colhead{RA}	&	
  \colhead{Dec} &
  \colhead{dist\tablenotemark{a}} &
  \colhead{P$_{false}$} &
  \colhead{SpT}&  
  \colhead{{\ewha}}&
  \colhead{$r-z$} &
  \colhead{$i-z$} &
  \colhead{{\lha}} &
  \colhead{Active?}
\\
  \colhead{}	&	
  \colhead{}	&	
  \colhead{(h m s)} &	
  \colhead{($^\circ~\arcmin~\arcsec$)}  &
  \colhead{(pc)}	&	  
  \colhead{(\%)} &
  \colhead{}&  
  \colhead{(\AA)} &
  \colhead{} &
  \colhead{} &
  \colhead{($\times10^{-4}$)} &
  \colhead{} 
}
\startdata
SLW 0149$+$2215\tablenotemark{b}	& A & 01 49 49.54 & $+$22 15 44.8 & 79	&  0.05 & M5 & 2.17	$\pm$ 0.30 &2.21 $\pm$ 0.02 & 0.77 $\pm$ 0.02 & 1.49  & y \\
		& B & 01 49 48.99 & $+$22 16 00.5 & 70	& & M4 & 4.56	$\pm$ 1.43 &2.46 $\pm$ 0.02 & 0.85 $\pm$ 0.02 &  2.81 & y 	\\
SLW 0214$-$1039	& A & 02 14 54.08 & $-$10 39 31.2 & 191	&0.53 & M4 & $-$0.14	$\pm$ 0.08 &1.89 $\pm$ 0.03 & 0.64 $\pm$ 0.03 & \nodata		& n	\\
		& B & 02 14 56.02 & $-$10 39 36.9 & 189	& & M4 & $-$0.10	$\pm$ 0.12 &1.94 $\pm$ 0.03 & 0.65 $\pm$ 0.03 & \nodata		& n 	\\
SLW 0248$-$0108	& A & 02 48 48.90 & $-$01 08 12.4 & 142	& 0.14 & M3 & 0.00	$\pm$ 0.07 &1.63 $\pm$ 0.03 & 0.55 $\pm$ 0.03 & \nodata		& n	\\
		& B & 02 48 50.02 & $-$01 08 17.7 & 139	& & M3 & $-$0.09	$\pm$ 0.11 &1.66 $\pm$ 0.03 & 0.56 $\pm$ 0.03 & \nodata		& n 	\\
SLW 0258$-$0711	& A & 02 58 07.59 & $-$07 11 06.5 & 236	& 0.83 & M4 & $-$0.24	$\pm$ 0.09 &1.98 $\pm$ 0.01 & 0.70 $\pm$ 0.01 & \nodata		& n 	\\
		& B & 02 58 04.73 & $-$07 10 42.7 & 255	& & M4 & $-$0.07	$\pm$ 0.12 & 1.99 $\pm$ 0.01 & 0.71 $\pm$ 0.01 & \nodata		& n 	\\
SLW 0421$+$0626	& A & 04 21 10.53 & $+$06 26 42.5 & 157	& 0.08 & M3 & 0.06	$\pm$ 0.06 &1.93 $\pm$ 0.01 & 0.66 $\pm$ 0.01 & \nodata		& n 	\\
		& B & 04 21 08.91 & $+$06 27 00.7 & 144	& & M3 & 0.04	$\pm$ 0.08 &2.11 $\pm$ 0.01 & 0.73 $\pm$ 0.01 &  \nodata		& n 	\\
SLW 0450$-$0412	& A & 04 50 10.52 & $-$04 12 33.5 & 49	& 0.01 & M3 & 0.14	$\pm$ 0.06 &2.07 $\pm$ 0.02 & 0.71 $\pm$ 0.02 & \nodata		& n 	\\
		& B & 04 50 09.71 & $-$04 12 29.6 & 51	& & M2 & $-$0.02	$\pm$ 0.10 &2.18 $\pm$ 0.02 & 0.74 $\pm$ 0.03 &  \nodata		& n 	\\
SLW 0741$+$1955\tablenotemark{b}	& A & 07 41 55.34 & $+$19 55 45.8 & 66	& 0.07 & M4 & 1.95	$\pm$ 0.57 &2.22 $\pm$ 0.01 & 0.79 $\pm$ 0.01 & 0.97 & y 	\\
		& B & 07 41 57.06 & $+$19 55 33.2 & 78	& & M4 & 5.25	$\pm$ 0.80 &2.46 $\pm$ 0.01 & 0.90 $\pm$ 0.01 &  2.88 & y 	\\
SLW 0756$+$4619	& A & 07 56 44.79 & $+$46 19 57.2 & 134	& 0.09 & M4 & $-$0.23	$\pm$ 0.06 &1.91 $\pm$ 0.02 & 0.64 $\pm$ 0.02 & \nodata		& n 	\\
		& B & 07 56 46.47 & $+$46 19 46.8 & 136	& & M4 & $-$0.19	$\pm$ 0.06 &1.93 $\pm$ 0.02 & 0.65 $\pm$ 0.02 &  \nodata		& n	\\
SLW 0819$+$3206	& A & 08 19 47.08 & $+$32 06 53.8 & 262	& 1.20 & M4 & $-$0.14	$\pm$ 0.15 &1.91 $\pm$ 0.01 & 0.67 $\pm$ 0.01 & \nodata		& n 	\\
		& B & 08 19 46.57 & $+$32 06 43.0 & 251	& & M3 & 0.07	$\pm$ 0.14 &1.94 $\pm$ 0.01 & 0.67 $\pm$ 0.01 &  \nodata		& n \\
SLW 0819$+$4731	& A & 08 19 21.00 & $+$47 31 59.3 & 205	& 0.57 & M2 & $-$0.24	$\pm$ 0.06 &1.29 $\pm$ 0.03 & 0.49 $\pm$ 0.03 & \nodata		& n 	\\
		& B & 08 19 26.10 & $+$47 32 14.1 & 211	& & M2 & $-$0.08	$\pm$ 0.07 &1.34 $\pm$ 0.03 & 0.49 $\pm$ 0.03 &  \nodata		& n 	\\
SLW 0835$+$3826	& A & 08 35 30.90 & $+$38 26 12.4 & 117	& 0.23& M3 & $-$0.15	$\pm$ 0.06 &1.84 $\pm$ 0.03 & 0.64 $\pm$ 0.03 & \nodata		& n	\\
		& B & 08 35 30.94 & $+$38 26 30.8 & 117	& & M3 & $-$0.14	$\pm$ 0.07 &1.88 $\pm$ 0.03 & 0.64 $\pm$ 0.03 &  \nodata		& n 	\\
SLW 0847$+$2539	& A & 08 47 48.94 & $+$25 39 29.7 & 169	& 0.36&  M3 & 0.01	$\pm$ 0.07 &1.62 $\pm$ 0.01 & 0.55 $\pm$ 0.01 &  \nodata		& n 	\\
		& B & 08 47 50.33 & $+$25 39 34.2 & 166	& & M3 & $-$0.05	$\pm$ 0.08 &1.66 $\pm$ 0.01 & 0.57 $\pm$ 0.01 &  \nodata		& n 	\\
SLW 0853$+$0325	& A & 08 53 58.87 & $+$03 25 07.9 & 185	& 0.29 &M3 & $-$0.20	$\pm$ 0.07 &1.62 $\pm$ 0.02 & 0.57 $\pm$ 0.03 & \nodata		& n \\
		& B & 08 54 01.31 & $+$03 25 27.0 & 191	& & M3 & 0.14	$\pm$ 0.07 &1.66 $\pm$ 0.02 & 0.58 $\pm$ 0.03 &  \nodata		& n \\
SLW 0858$+$0936\tablenotemark{b}	& A & 08 58 57.80 & $+$09 36 59.1 & 65	& 0.01 & M4 & 2.66	$\pm$ 0.39 &2.12 $\pm$ 0.02 & 0.72 $\pm$ 0.02 & 1.67 & y 	\\
		& B & 08 58 54.73 & $+$09 37 23.7 & 63	& & M4 & 4.90	$\pm$ 2.07 &2.20 $\pm$ 0.02 & 0.75 $\pm$ 0.02 &  2.71 & y 	\\
SLW 0919$+$2417	& A & 09 19 41.98 & $+$24 17 10.5 & 173	& 0.41 & M5 & 0.36	$\pm$ 0.11 &2.25 $\pm$ 0.02 & 0.77 $\pm$ 0.02 & \nodata		& n 	\\
		& B & 09 19 42.49 & $+$24 17 24.2 & 166	& & M5 & $-$0.05	$\pm$ 0.27 &2.26 $\pm$ 0.02 & 0.79 $\pm$ 0.02 &  \nodata		& n 	\\
SLW 0934$+$1512	& A & 09 34 35.01 & $+$15 12 41.8 & 128	& 0.13 & M4 & $-$0.41	$\pm$ 0.06 &2.05 $\pm$ 0.02 & 0.70 $\pm$ 0.02 & \nodata		& n	\\
		& B & 09 34 33.16 & $+$15 12 58.9 & 138	& & M4 & $-$0.39	$\pm$ 0.08 &2.09 $\pm$ 0.02 & 0.73 $\pm$ 0.02 &  \nodata		& n 	\\
SLW 0949$+$0404	& A & 09 49 17.24 & $+$04 04 15.7 & 194	& 0.24 & M4 & $-$0.38	$\pm$ 0.13 &2.04 $\pm$ 0.03 & 0.70 $\pm$ 0.02 & \nodata		& n 	\\
		& B & 09 49 15.94 & $+$04 03 53.1 & 191	& & M4 & $-$0.19	$\pm$ 0.09 &2.06 $\pm$ 0.03 & 0.71 $\pm$ 0.02 &  \nodata		& n 	\\
SLW 0951$+$3709	& A & 09 51 26.44 & $+$37 09 55.0 & 144	& 0.32 & M5 & 2.82	$\pm$ 0.29 &2.36 $\pm$ 0.03 & 0.84 $\pm$ 0.03 & 1.83 & y 	\\
		& B & 09 51 24.91 & $+$37 10 35.0 & 145	& & M5 & 1.89	$\pm$ 0.27 &2.40 $\pm$ 0.03 & 0.83 $\pm$ 0.03 &  1.23 & y 	\\
SLW 0954$+$0647	& A & 09 54 06.37 & $+$06 47 03.7 & 114	& 0.07 & M3 & 2.37	$\pm$ 0.08 &1.82 $\pm$ 0.02 & 0.65 $\pm$ 0.02 & 1.54 & y 	\\
		& B & 09 54 02.54 & $+$06 45 49.7 & 111	& & M3 & 5.49	$\pm$ 0.09 &1.90 $\pm$ 0.02 & 0.68 $\pm$ 0.02 & 3.56 & y 	\\
SLW 1022$+$1733	& A & 10 22 50.92 & $+$17 33 14.3 & 203	& 0.35 & M3 & 0.52	$\pm$ 0.12 & 1.71 $\pm$ 0.03 & 0.58 $\pm$ 0.03 & 0.422 & w \\
		& B & 10 22 51.50 & $+$17 33 07.8 & 209	& & M3 & $-$0.15	$\pm$ 0.11 &1.72 $\pm$ 0.03 & 0.58 $\pm$ 0.03 &  \nodata		& n 	\\
SLW 1032$+$3823	& A & 10 32 00.11 & $+$38 23 33.4 & 152	& 0.23 & M3 & $-$0.22	$\pm$ 0.07 & 2.10 $\pm$ 0.02 & 0.76 $\pm$ 0.03 & \nodata		& n 	\\
		& B & 10 31 59.30 & $+$38 23 37.1 & 138	& & M4 & $-$0.12	$\pm$ 0.12 &2.21 $\pm$ 0.02 & 0.81 $\pm$ 0.03 & \nodata		& n 	\\
SLW 1034$+$4040	& A & 10 34 15.27 & $+$40 40 58.9 & 58	& 0.00 & M3 & $-$0.65	$\pm$ 0.08 & 2.24 $\pm$ 0.03 & 0.79 $\pm$ 0.03 & \nodata		& n 	\\
		& B & 10 34 09.33 & $+$40 39 39.6 & 50	& & M3 & $-$0.20	$\pm$ 0.10 &2.35 $\pm$ 0.03 & 0.85 $\pm$ 0.03 &  \nodata		& n\\
SLW 1036$+$0118	& A & 10 36 49.54 & $+$01 18 51.2 & 147	& 0.12 & M0 & $-$0.30	$\pm$ 0.10 & 1.14 $\pm$ 0.02 & 0.40 $\pm$ 0.03 & \nodata		& n 	\\
		& B & 10 36 50.17 & $+$01 19 04.8 & 155	& & M3 & 0.23	$\pm$ 0.37 &1.78 $\pm$ 0.02 & 0.62 $\pm$ 0.03 & \nodata		& n 	\\
SLW 1120$+$2046\tablenotemark{b}	& A & 11 20 03.38 & $+$20 46 53.2 & 96	& 0.06 & M4 & 1.65	$\pm$ 0.74 &2.13 $\pm$ 0.03 & 0.72 $\pm$ 0.03 & 1.60 & y	\\
		& B & 11 20 05.26 & $+$20 46 54.8 & 101	& & M4 & 5.49	$\pm$ 0.91 &2.32 $\pm$ 0.03 & 0.80 $\pm$ 0.03 & 2.52 & y	\\
SLW 1133$+$0035	& A & 11 33 36.99 & $+$00 35 20.5 & 71	&0.01 & M4 & $-$0.64	$\pm$ 0.08 & 2.23 $\pm$ 0.02 & 0.80 $\pm$ 0.01 & \nodata		& n	\\
		& B & 11 33 37.48 & $+$00 35 14.6 & 70	& & M4 & $-$0.43	$\pm$ 0.09 &2.30 $\pm$ 0.02 & 0.83 $\pm$ 0.01 &  \nodata		& n 	\\
SLW 1204$+$1951	& A & 12 04 36.35 & $+$19 51 34.6 & 66	&0.02 & M5 & $-$0.47	$\pm$ 0.06 &2.72 $\pm$ 0.01 & 0.95 $\pm$ 0.02 & \nodata		& n 	\\
		& B & 12 04 35.97 & $+$19 51 47.6 & 75	& & M5 & $-$0.75	$\pm$ 0.10 &2.89 $\pm$ 0.01 & 1.01 $\pm$ 0.02 & \nodata		& n	\\
SLW 1239$-$0305	& A & 12 39 21.63 & $-$03 05 57.3 & 102	& 0.12 & M4 & 1.32	$\pm$ 0.26 & 2.41 $\pm$ 0.02 & 0.88 $\pm$ 0.01 & 0.857 & y 	\\
		& B & 12 39 20.76 & $-$03 07 30.5 & 113	& & M4 & 3.14	$\pm$ 0.26 &2.48 $\pm$ 0.01 & 0.91 $\pm$ 0.01 & 2.04 & y 	\\
SLW 1241$+$4645	& A & 12 41 00.45 & $+$46 45 29.1 & 267	& 1.48 &M0 & $-$0.43	$\pm$ 0.06 &1.07 $\pm$ 0.02 & 0.37 $\pm$ 0.02 & \nodata		& n 	\\
		& B & 12 40 59.75 & $+$46 45 09.9 & 294	& & M2 & $-$0.15	$\pm$ 0.12 & 1.42 $\pm$ 0.02 & 0.50 $\pm$ 0.02 &  \nodata		& n 	\\
SLW 1348$+$0811	& A & 13 48 02.47 & $+$08 11 40.2 & 125	& 0.16 & M2 & $-$0.22	$\pm$ 0.08 &1.76 $\pm$ 0.02 & 0.63 $\pm$ 0.02 & \nodata		& n \\
		& B & 13 48 00.48 & $+$08 12 31.1 & 121	& & M3 & 0.11	$\pm$ 0.10 &1.92 $\pm$ 0.02 & 0.67 $\pm$ 0.02 &  \nodata		& n	\\
SLW 1408$+$1130	& A & 14 08 54.13 & $+$11 30 45.0 & 133	& 0.15 & M3 & $-$0.28	$\pm$ 0.07 & 1.91 $\pm$ 0.03 & 0.63 $\pm$ 0.03 & \nodata		& n 	\\
		& B & 14 08 49.87 & $+$11 30 22.3 & 146	& & M3 & $-$0.22	$\pm$ 0.14 &2.09 $\pm$ 0.03 & 0.71 $\pm$ 0.03 &  \nodata		& n 	\\
SLW 1441$+$0156	& A & 14 41 26.23 & $+$01 56 02.0 & 94	&0.31 & M4 & $-$0.40	$\pm$ 0.06 &2.29 $\pm$ 0.03 & 0.81 $\pm$ 0.03 & \nodata		& n	\\
		& B & 14 41 27.55 & $+$01 56 16.4 & 80	& & M4 & $-$0.33	$\pm$ 0.10 &2.49 $\pm$ 0.03 & 0.89 $\pm$ 0.03 &  \nodata		& n \\
SLW 1446$+$5324	& A & 14 46 55.22 & $+$53 24 22.5 & 86	& 0.13 & M4 & $-$0.16	$\pm$ 0.07 &2.17 $\pm$ 0.03 & 0.78 $\pm$ 0.03 & \nodata		& n 	\\
		& B & 14 46 55.57 & $+$53 25 17.3 & 96	& & M4 & 0.87	$\pm$ 0.14 &2.33 $\pm$ 0.03 & 0.84 $\pm$ 0.03 &  0.195 & w 	\\
SLW 1502$+$6057	& A & 15 02 01.04 & $+$60 57 11.0 & 151	& 0.34 & M2 & 0.01	$\pm$ 0.06 &1.52 $\pm$ 0.02 & 0.52 $\pm$ 0.02 & \nodata		& n	\\
		& B & 15 01 59.85 & $+$60 57 36.3 & 151	& & M2 & $-$0.14	$\pm$ 0.07 & 1.53 $\pm$ 0.02 & 0.52 $\pm$ 0.02 &  \nodata		& n 	\\
SLW 1545$+$1859	& A & 15 45 35.39 & $+$18 59 29.4 & 75	& 0.08 & M4 & $-$0.10	$\pm$ 0.05 &2.48 $\pm$ 0.03 & 0.90 $\pm$ 0.02 & \nodata		& n 	\\
		& B & 15 45 34.62 & $+$18 59 33.1 & 85	& & M5 & 0.06	$\pm$ 0.08 &2.67 $\pm$ 0.03 & 0.98 $\pm$ 0.02 &  \nodata		& n	\\
SLW 1546$+$3928	& A & 15 46 54.08 & $+$39 28 17.0 & 202	& 0.83 & M2 & 0.03	$\pm$ 0.16 &1.65 $\pm$ 0.03 & 0.64 $\pm$ 0.03 & \nodata		& n 	\\
		& B & 15 46 55.89 & $+$39 28 07.8 & 229	& & M3 & $-$0.12	$\pm$ 0.25 & 1.76 $\pm$ 0.03 & 0.67 $\pm$ 0.03 &  \nodata		& n 	\\
SLW 1554$+$4425	& A & 15 54 04.02 & $+$44 25 44.1 & 143	& 0.16 & M2 & $-$0.20	$\pm$ 0.06 &1.61 $\pm$ 0.03 & 0.52 $\pm$ 0.04 & \nodata		& n 	\\
		& B & 15 54 03.81 & $+$44 25 30.2 & 147	& & M3 & $-$0.23	$\pm$ 0.09 &1.66 $\pm$ 0.03 & 0.53 $\pm$ 0.04 & \nodata		& n 	\\
SLW 1914$+$7928	& A & 19 14 45.31 & $+$79 28 49.5 & 305	& 4.15 & M2 & 0.16	$\pm$ 0.05 &1.51 $\pm$ 0.02 & 0.51 $\pm$ 0.03 & \nodata		& n 	\\
    		& B & 19 14 48.26 & $+$79 28 38.1 & 331	& & M2 & $-$0.18	$\pm$ 0.07 &1.63 $\pm$ 0.02 & 0.55 $\pm$ 0.03 &  \nodata		& n 	\\
SLW 2112$-$0044 & A & 21 12 56.83 & $-$00 44 45.4 & 324	& 1.03 & M1 & $-$0.15	$\pm$ 0.07 &1.27 $\pm$ 0.01 & 0.47 $\pm$ 0.01 & \nodata		& n 	\\
		& B & 21 12 56.35 & $-$00 44 57.5 & 323	& & M1 & 0.00	$\pm$ 0.07 & 1.28 $\pm$ 0.01 & 0.47 $\pm$ 0.01 &  \nodata		& n \\
SLW 2208$-$0000 & A & 22 08 22.50 & $-$00 00 33.3 & 163	& 0.56 & M4 & 0.18	$\pm$ 0.06 & 2.06 $\pm$ 0.02 & 0.73 $\pm$ 0.03 & \nodata		& n \\
		& B & 22 08 23.00 & $-$00 00 26.9 & 139	& & M4 & 0.20	$\pm$ 0.10 &2.12 $\pm$ 0.02 & 0.74 $\pm$ 0.03 &  \nodata		& n 	\\
SLW 2212$+$4015 & A & 22 12 31.82 & $+$40 15 23.3 & 147	& 0.35 & M4 & $-$0.34	$\pm$ 0.07 &2.20 $\pm$ 0.01 & 0.77 $\pm$ 0.01 & \nodata		& n 	\\
		& B & 22 12 34.81 & $+$40 15 08.8 & 139	& & M4 & $-$0.29	$\pm$ 0.08 &2.26 $\pm$ 0.01 & 0.79 $\pm$ 0.01 & \nodata		& n 	\\
SLW 2244$+$2318 & A & 22 44 20.21 & $+$23 18 04.4 & 91	& 0.06 &  M4 & $-$0.42	$\pm$ 0.07 &2.29 $\pm$ 0.02 & 0.82 $\pm$ 0.02 & \nodata		& n 	\\
		& B & 22 44 19.96 & $+$23 18 13.6 & 84	& & M4 & $-$0.47	$\pm$ 0.08 &2.31 $\pm$ 0.02 & 0.84 $\pm$ 0.02 &  \nodata		& n 	\\
SLW 2253$+$2726 & A & 22 53 03.55 & $+$27 26 46.3 & 70	& 0.55 & M5 & 0.79	$\pm$ 0.05 & 2.49 $\pm$ 0.02 & 0.87 $\pm$ 0.03 & 0.513  & w 	\\
		& B & 22 53 04.31 & $+$27 26 54.1 & 67	& & M5 & 0.30	$\pm$ 0.08 & 2.67 $\pm$ 0.02 & 0.94 $\pm$ 0.03 &  \nodata		& n 	\\
SLW 2254$-$0931 & A & 22 54 51.34 & $-$09 31 29.5 & 112	& 0.22 & M4 & $-$0.25	$\pm$ 0.06 &2.18 $\pm$ 0.02 & 0.75 $\pm$ 0.02 & \nodata		& n 	\\
		& B & 22 54 53.89 & $-$09 32 36.7 & 128	& & M5 & $-$0.40	$\pm$ 0.12 &2.32 $\pm$ 0.03 & 0.80 $\pm$ 0.04 & \nodata		& n 	\\
SLW 2258$+$2806 & A & 22 58 58.20 & $+$28 06 25.2 & 211	& 0.61 & M5 & 0.04	$\pm$ 0.05 &1.87 $\pm$ 0.02 & 0.64 $\pm$ 0.03 & \nodata		& n 	\\
		& B & 22 58 56.47 & $+$28 05 54.8 & 203	& & M4 & 0.20	$\pm$ 0.05 & 1.91 $\pm$ 0.02 & 0.66 $\pm$ 0.03 &  \nodata		& n \\
SLW 2309$+$2440 & A & 23 09 32.55 & $+$24 40 25.9 & 177	& 0.42 & M2 & $-$0.07	$\pm$ 0.08 &1.53 $\pm$ 0.03 & 0.55 $\pm$ 0.02 & \nodata		& n \\
		& B & 23 09 32.26 & $+$24 40 48.9 & 182	& & M2 & 0.08	$\pm$ 0.08 & 1.60 $\pm$ 0.03 & 0.56 $\pm$ 0.02 & \nodata		& n 	\\
SLW 2310$+$2201 & A & 23 10 50.47 & $+$22 01 24.6 & 175	& 0.31 & M3 & $-$0.15	$\pm$ 0.07 &1.83 $\pm$ 0.02 & 0.64 $\pm$ 0.03 & \nodata		& n 	\\
		& B & 23 10 51.90 & $+$21 59 50.7 & 180	& & M3 & $-$0.10	$\pm$ 0.09 & 1.95 $\pm$ 0.02 & 0.68 $\pm$ 0.03 & \nodata		& n 	\\
SLW 2311$+$0802 & A & 23 11 24.12 & $+$08 02 51.5 & 154	&0.24 &  M5 & 0.32	$\pm$ 0.09 &2.14 $\pm$ 0.02 & 0.74 $\pm$ 0.03 & \nodata		& n 	\\
		& B & 23 11 23.53 & $+$08 02 52.2 & 146	& & M5 & 0.35	$\pm$ 0.10 &2.20 $\pm$ 0.02 & 0.76 $\pm$ 0.03 &  \nodata		& n 	\\
SLW 2315$-$0045\tablenotemark{b} & A & 23 15 44.02 & $-$00 45 00.6 & 50	& 0.01 & M5 & 3.96	$\pm$ 0.31 &2.63 $\pm$ 0.03 & 0.92 $\pm$ 0.03 & 2.12  & y 	\\
		& B & 23 15 46.63 & $-$00 44 06.2 & 51	& & M5 & 3.01	$\pm$ 0.26 & 2.79 $\pm$ 0.03 & 0.98 $\pm$ 0.03 & 1.91  & y 	
\enddata
\tablenotetext{a}{Distances were calculated using the \citet{Bochanski2010} $M_r$ as a function of $r-z$.}
\tablenotetext{b}{For binaries with time-resolved observations, the mean and standard deviation of observations, with flares included, are listed for H$\alpha$ EW.}
\end{deluxetable*}

\end{document}